\documentclass{PoS}
\usepackage{subfig}

\title{Astrophysical Tau Neutrino Identification with IceCube Waveforms}

\ShortTitle{Astrophysical Tau Neutrinos in IceCube}

\author{
The IceCube Collaboration\footnote{For collaboration list, see PoS(ICRC2019) 1177.}\\
{\itshape \href{http://icecube.wisc.edu/collaboration/authors/icrc19_icecube}{http://icecube.wisc.edu/collaboration/authors/icrc19\_icecube}}\\
E-mail: \email{lwille@icecube.wisc.edu, donglianxu@sjtu.edu.cn}
}

\abstract{

The standard neutrino oscillation paradigm predicts almost equal fractions of astrophysical neutrino flavors at Earth regardless of their production ratio at the sources. Therefore, identification of astrophysical tau neutrinos could not only reconfirm the astrophysical neutrino flux measured by IceCube, but also is essential in precisely determining the astrophysical neutrino flavor ratio at Earth, which is an important probe for physics beyond the Standard Model over astronomical baselines. A tau neutrino undergoing a charged current (CC) interaction in IceCube could produce a double deposition of energy, with the first one from the CC hadronic shower and the second from the subsequent tau lepton decay shower. Above an energy of ~100 TeV, such consecutive energy depositions might be resolvable in the sensor waveforms and hence can be a signature of an individual tau neutrino interaction in IceCube. We will present the results of a search for astrophysical tau neutrinos in IceCube waveforms with improved double pulse waveform identification techniques and using 8 years of data. \\

\vspace{4mm}
{\bfseries Corresponding authors:}
\speaker{Logan Wille} and Donglian Xu\footnote{now at Tsung-Dao Lee Institute}\\
{\itshape Wisconsin IceCube Particle Astrophysics Center, University of Wisconsin - Madison}\\

}

\FullConference{36th International Cosmic Ray Conference -ICRC2019-\\
		July 24th - August 1st, 2019\\
		Madison, WI, U.S.A.}

\begin{document}

\section{Introduction}\label{sec:info}
With the discovery of the diffuse astrophysical neutrinos by the IceCube Neutrino Observatory, neutrino oscillations can be explored at new energy and distance scales \cite{HESE} \cite{global} \cite{aachen}. Tau neutrinos are rarely produced in nature, however there may be a flux of astrophysical tau neutrinos arising from the oscillation of the other astrophysical neutrinos. Measuring astrophysical tau neutrinos can allow us to study the neutrino oscillations over very long baselines and at high energies never before explored \cite{carlos}. This analysis works towards identifying tau neutrinos interacting in IceCube through a unique detection channel that is complementary to a machine learning based analysis \cite{icrcMLDP}. Previous waveform-based analyses have searched for tau neutrinos in IceCube data but did not observe any tau neutrinos \cite{2017icrcDC} \cite{PRD}.

IceCube is a cubic-kilometer neutrino detector installed at the geographic South Pole between ice depths of 1450 m and 2450 m, and was completed in 2010 \cite{Aartsen:2016nxy}. Reconstruction of the direction, energy and flavor of the neutrinos relies on the detection of Cherenkov radiation emitted by charged particles produced in the interactions of neutrinos in the surrounding ice or the nearby bedrock.

\section{Tau Neutrinos in IceCube}

Tau neutrinos of sufficiently high energy undergoing charged-current interactions have a distinct topology in IceCube, called a "double bang" which produces two large energy losses, the first from the initial hadronic interaction with a nuclei of an ice molecule and the second from the tau lepton decaying through hadrons or electrons. Tau leptons have a low interaction cross section and can travel freely through the ice without losing energy, but since their lifetime is very short they decay rapidly, not far from their creation point. These traits of the tau lepton make the tau neutrino interaction unique, no other neutrino produces such a signature.

The flux of astrophysical tau neutrinos follows a falling spectrum, meaning there are considerably more low energy neutrinos than high energy neutrinos. This results in far more low energy events and thus, tau neutrino events with tau leptons that travel on the order of tens of meters are far more common than hundreds of meters. When a tau lepton travels only a few tens of meters IceCube cannot easily resolve the position of the two energy depositions due to the sparse nature of the detector, the event will look like a single cascade. However, a single DOM can easily detect the light arriving from the initial interaction and subsequent tau lepton decay, which can appeare as two bumps in the waveform, as shown in Fig.~\ref{fig:doublepulsewaveform}. This type of waveform is referred to as a double pulse, and is used by this analysis to identify tau neutrino events.

\section{Event Selection}

In order to observe a tau neutrino in IceCube data, significant cuts are needed to reduce the background events, non-double pulse tau neutrinos, electron neutrinos, muon neutrinos, and atmospheric muons. These backgrounds are targeted in three different steps. The first level is a double pulse algorithm (DPA), which focuses on selecting double pulse tau neutrinos and rejecting single cascade events and tau neutrinos that do not produce a double pulse. The second is a topology cut, where cascade-like events are selected and longer, track-like events created by muons and muon neutrinos are rejected. Finally, a containment cut that selects only events that start inside the detector, which excludes the most remaining atmospheric muons and brings the atmospheric backgrounds to sub-dominant level. 

\begin{figure}[!h]
   \centering
   \includegraphics[width=0.60\textwidth]{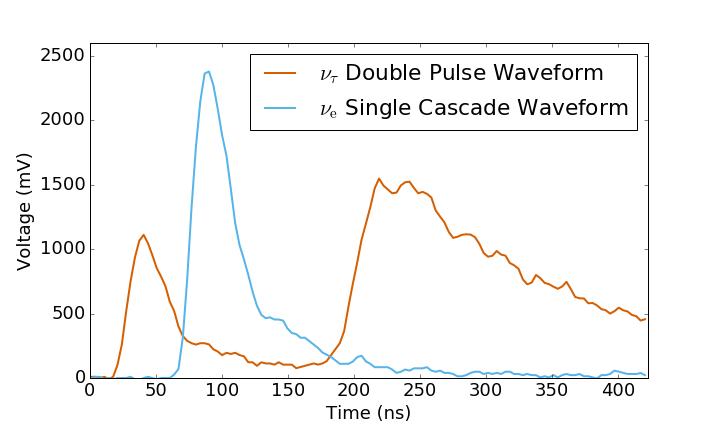}
    \caption{A double pulse waveform from a tau neutrino charged-current interaction is shown, indicating the first pulse's rising and falling edges along with the second pulse's rising edge. For comparison, a single pulse from an electron neutrino interaction with only one pulse is also shown.}
    \label{fig:doublepulsewaveform}
\end{figure}

 The necessary features for DPA to identify a double pulse are the rising and falling edge of the first pulse and the rising edge of the second pulse. A second pulse falling edge is not necessary to search for as it is a guaranteed feature and offers no discriminating power. An example double pulse waveform is shown in Fig.~\ref{fig:doublepulsewaveform}. The DPA and its enhancement by incorperating neighboring DOM waveforms into the algorithm is detailed in Ref. \cite{icrc}.

The remaining backgrounds are atmospheric muons and muon neutrinos. Muons are long lived compared to the tau lepton, and so can travel several kilometers through the ice. These long muons create events, called tracks, that extend through the detector which are distinct from the very localized tau neutrino events, called cascades. To classify an event as either cascade or track two reconstructions are applied, one cascade and one track. The likelihood value of the resulting cascade and track best fits are a measure of how well each the event topology fits with the two different assumed event types. To combine these two reconstructions and determine which of the two topologies an event most likely falls under, the difference between the reduced log likelihood of the reconstructions is taken. A cascade will have a large likelihood for the cascade reconstruction and small likelihood for the track reconstruction and vice-versa for a track.

\begin{figure}[!h]
   \centering
   \includegraphics[width=1.\textwidth]{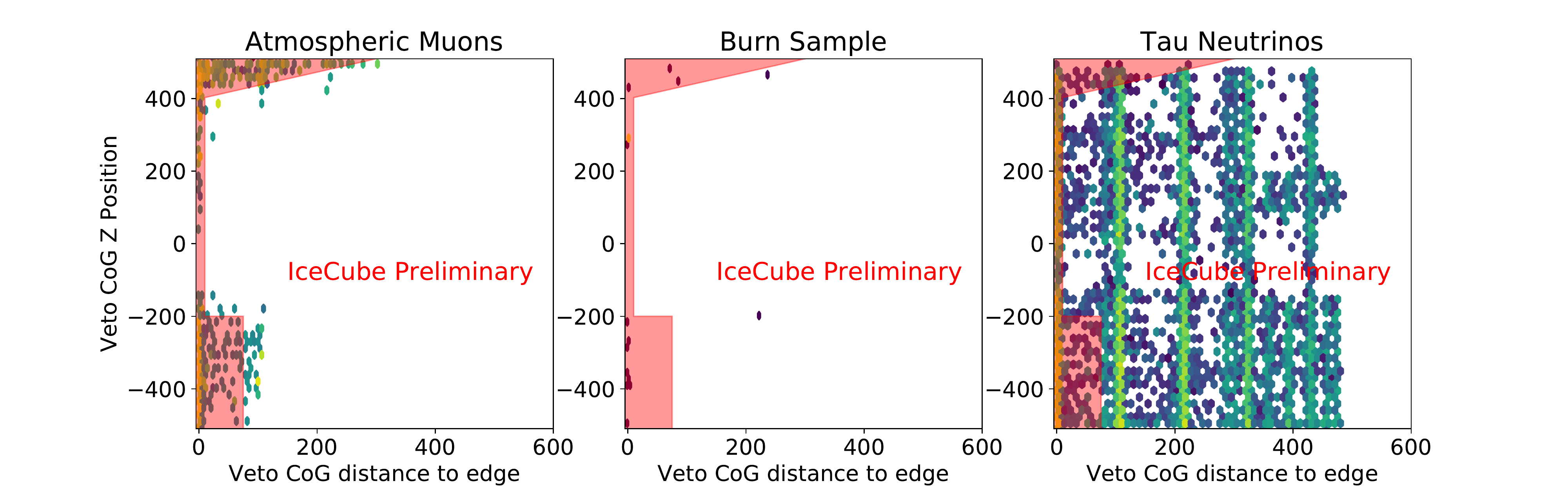}
    \caption{The distribution of events in the detector using the $\mathbf{R}_{250 PE}$ CoG position. The atmospheric muons cluster near the edge, top and bottom corners, while the selected tau neutrinos tend to cluster near the string layers across the whole detector. The red shaded area are the rejected regions.}
    \label{fig:veto}
\end{figure}

In addition to the topology cut, a geometric containment cut is employed to remove atmospheric muons. The containment cut requires the events to start inside the IceCube detector and not near the edge, which rejects the many remaining muon events while keeping a large fraction of tau neutrino events. The containment cut uses a single variable, $\mathbf{R}_{250 PE}$, the center of gravity of the first 250 photoelectrons,

$\mathbf{R}_{250 PE} = \frac{1}{250 PE}\sum_{i=1} \mathbf{r_i}*C_i$,

where $r_i$ is the position of the $i$th DOM and $C_i$ is the charge that DOM observed until 250 PE total charge was observed in an event. Thus $\mathbf{R}_{250 PE}$ estimates where the very beginning of an event is, and shows if an event was starting inside the detector or near an edge. The three dimensional center of gravity is distilled into two variables, distance to closest edge, $E_{veto}$, and height, $Z_{veto}$, in the detector. The distribution of these two variables for the muon background, burn sample (10\% of total analyzed data), and signal tau neutrino events are shown in Fig. \ref{fig:veto}. The red shaded areas denote the three regions that are removed by the containment cut, the top corner, bottom corner, and edge of the detector.

The selection process has created a sample of events where tau neutrinos are the dominant event type. Table~\ref{tab:eventrates} summarizes the event rates at final cut level -- for 8 years of livetime this selection expects to find a total of 3.13 events, 1.71 of which are signal tau neutrinos.

\begin{table}
  \centering
  \begin{tabular}{| l | r |}
  	  \hline
       & Event Rate in 8 years \\ \hline
      NuTau CC &  1.72 $\pm$ 0.023 \\ \hline
      NuMu Astro. + Atmo. & 0.95 $\pm$ 0.048  \\ \hline
      NuE Astro. + Atmo. & 0.26 $\pm$ 0.010  \\ \hline
      Atmospheric Muons &  0.2 $\pm$ 0.14  \\ \hline
  \end{tabular}
  \caption{The final sample expected rate of events in the 8 years of data.
  \label{tab:eventrates}}
\end{table}

\section{Analysis Methods}

With an established event selection the relevant physical information can be extracted by using a binned maximum likelihood method (forward folding). In addition confidence intervals of these physical parameters are created using a Feldman-Cousins scan which ensures proper coverage for limited-statistics folding. Two observables are used to bin the data, the reconstructed energy ($E_{reco}$) and the maximum duration of the first pulse ($\Delta T_{max}$) for any waveform that passes the DPA. Due to limited number of expected events, the parameters are allowed to float freely except for the astrophysical spectra index, which is fixed to three previously measured values, -2.19, -2.5, and -2.9, from IceCube analyses \cite{aachen} \cite{global} \cite{HESE}. The binning used is the same as shown in Fig. \ref{fig:scores}.

In addition, a background p-value score is calculated for each event observed. To create the p-value, a test statistic (TS) value is found for each background event to create a TS distribution. For each observed event, the same TS value is calculated and compared to the background distribution to find the p-value for that event. For this analysis the event TS value is, 

\begin{equation}
\mathrm{TS} = \mathrm{Log}(\mathcal{L}(\lambda)/\mathcal{L}(\lambda = 0)). 
\label{eq:TS}
\end{equation}
where $\mathcal{L}$ is the per bin likelihood with a fitted parameter $\lambda$. The likelihood is maximized by fitting $\lambda$ for the bin an event lies in. The likelihood is defined as,
\begin{equation}
\mathcal{L}(\lambda) = (\mathrm{P}_B^{i,j} + \lambda \times \mathrm{P}_S^{i,j})/(\lambda +1), 
\label{eq:TSlikelihood}
\end{equation}
where $\mathrm{P}^{i,j}_{B,S}$ are the background, signal probability density function for the i,jth bin in ($E_{reco}$, $\Delta T_{max}$) phase space the event lies in, normalized as
\begin{equation}
\mathrm{P}_{B,S}^{i,j} = \frac{\mathrm{N}_{B,S}^{i,j}}{ \int \int \mathrm{N}_{B,S} \mathrm{dE dt}},
\label{eq:TSprob}
\end{equation}
$\mathrm{N}_{B,S}^{i,j}$ is the number of background, signal events observed in the i,jth bin which is dependent on the assumed astrophysical flux. The bins are set to the same as forward folding.

\section{Results}

\begin{figure}[!h]
   \centering
   \includegraphics[width=0.75\textwidth]{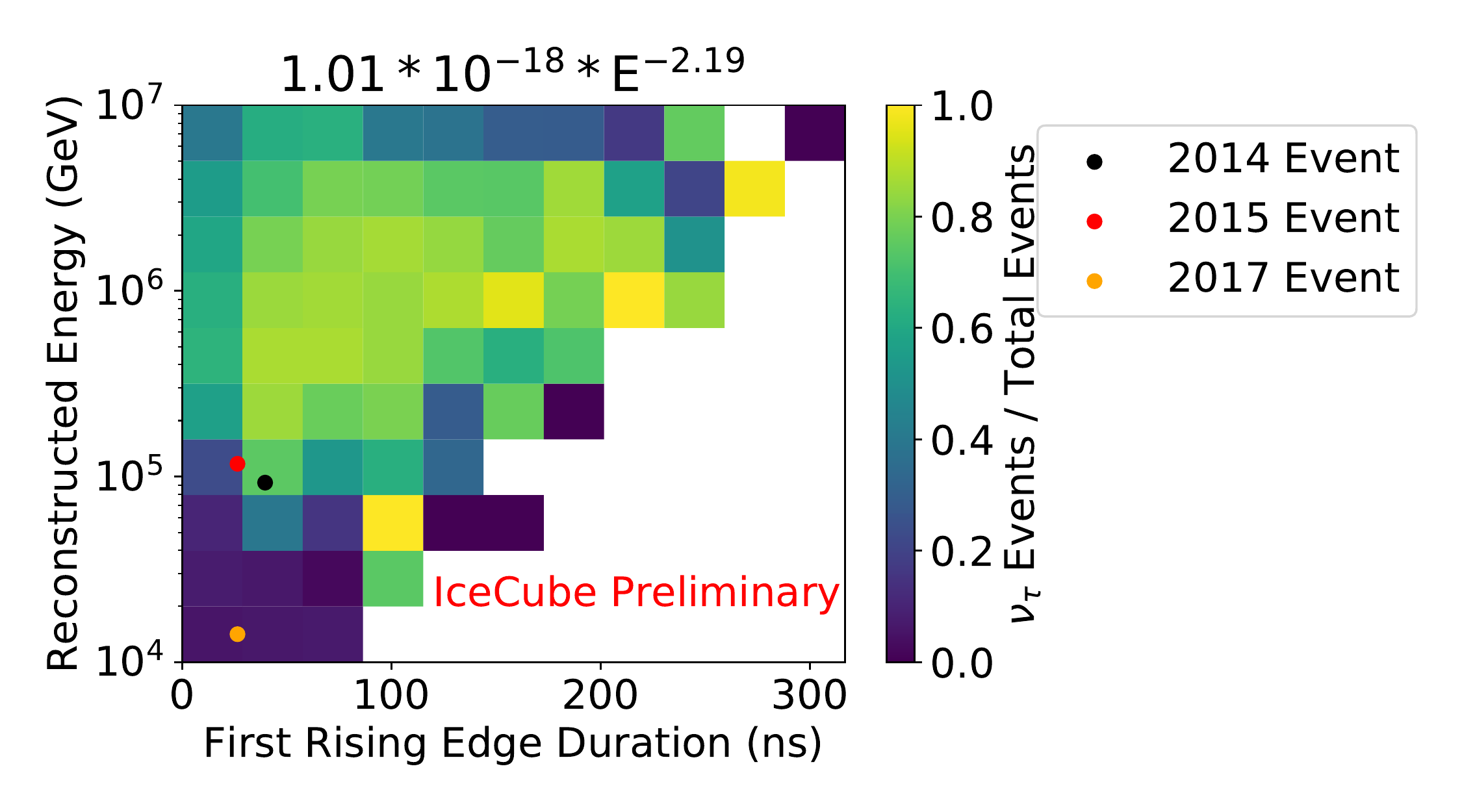}
    \caption{The signal to total event ratio binned in the two observables used for the forward folding and p-value. Also shown are where the three observed events lie in this phase space.}
    \label{fig:scores}
\end{figure}

The event selection was applied to 2759.66 cumulative days of data in a period between May 2010 and December 2018. In this data set a total of three events were observed, one in 2014, 2015, and 2017. The three events in relation to the expected signal to total events is shown in Fig.~\ref{fig:scores}. 

\begin{figure}[!h]
  \centering
  \subfloat{\includegraphics[width=0.44\textwidth]{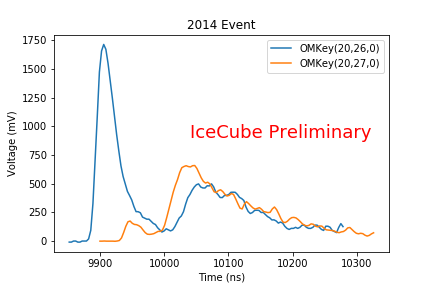}}
  \hfill
  \subfloat{\includegraphics[width=0.55\textwidth]{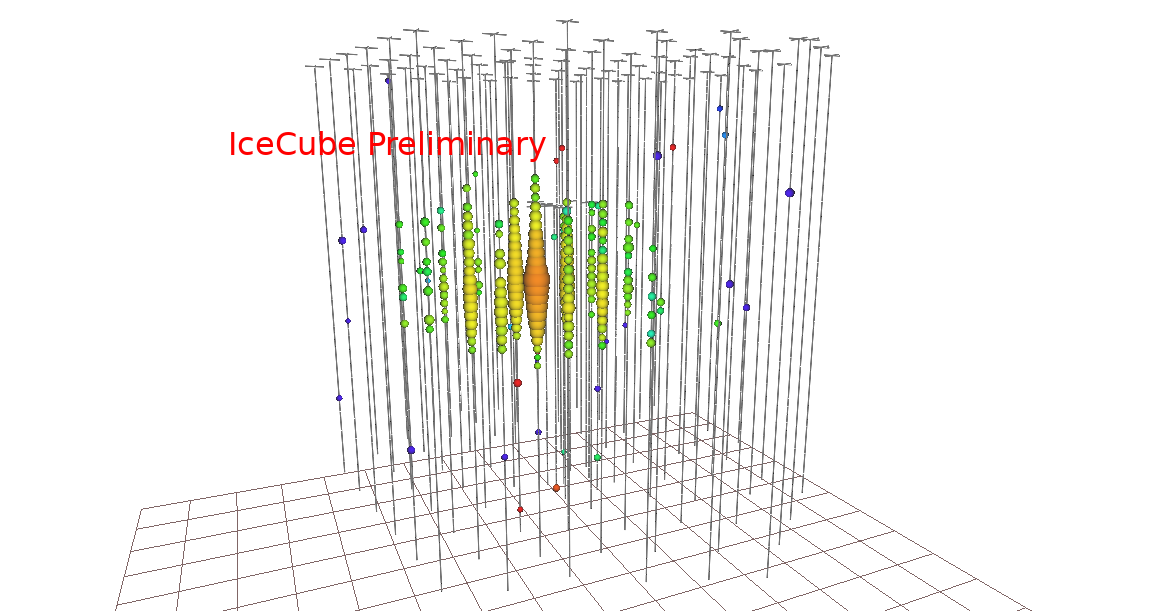}}
  \caption{\textbf{Left}: The two waveforms selected by the double pulse algoritm are shown. These waveforms are on neighboring DOMs (string and PMT ID numebrs are noted as OMKeys) and so pass the new local coincidence double pulse selection. The waveform recorded on DOM 20, 27 also passed the previous single DOM double pulse algorithm. \textbf{Right}: Event view of the 2014 event. The interaction appears to start inside of the detector volume, so a neutrino event is highly likely.}
  \label{fig:2014event}
\end{figure}

The event observed in 2014 is the most interesting event of the sample from the perspective of it having the highest probability of being a signal tau neutrino. As shown in Fig.~\ref{fig:2014event}, two waveforms passed the DPA, the event occurred in the middle of the detector, suggesting this event is likely a neutrino. Due to the lower reconstructed energy of 93 TeV, the p-values of this event are not significant, 0.29, 0.196, and 1.0 for $E^{-2.19}, E^{-2.5}$ and $E^{-2.9}$ respectively. A note of interest for this event, two other IceCube tau neutrino analyses one that uses machine learning and other that uses a reconstruction method to identify double cascade events also found this event as a tau neutrino candidate \cite{icrcMLDP} \cite{icrcDC}. Further work is under way to address the source of this event, the two analyses finding this event hints towards this event being a tau neutrino as both analyses have different dominant backgrounds.

\begin{figure}[!h]
  \centering
  \subfloat{\includegraphics[width=0.44\textwidth]{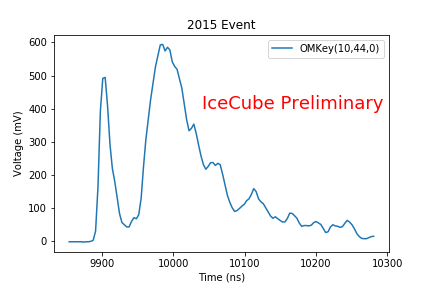}}
  \hfill
  \subfloat{\includegraphics[width=0.55\textwidth]{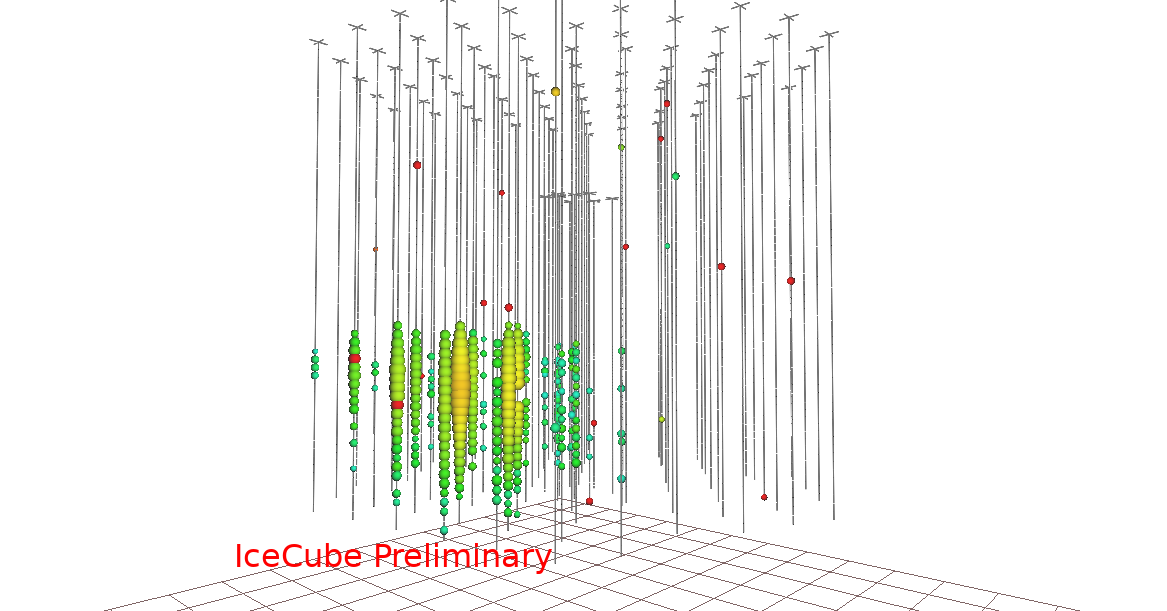}}
  \caption{\textbf{Left}: The double pulse waveform recorded for this event. Only one DOM recorded a double pulse, the neighboring DOMs have no evident double pulse signature. The first pulse is sharp with a short duration first rising edge and falling edge, suggestive of a muon cherenkov light producing the first pulse. \textbf{Right}: The event view of the 2015 event. This event starts inside of the detector going in a horizontal direction. There are a few hits on the left side of the detector that hint towards this event containing a muon that leaves the detector.}
  \label{fig:2015event}
\end{figure}

The event observed in 2015 is a less significant event with a p-value of 1.0 for all three assumed spectra. This event passed only the single DOM DPA configuration, the waveform that passed the cut is shown in Fig.~\ref{fig:2015event}. One of the characteristics that shows this event is background like is the sharp, short duration first pulse, which is a typical signature of Cherenkov radiation from a passing muon. The reconstructed energy is 117 TeV, once again reducing the likelihood for this event being a signal tau neutrino as most of the expected signal events are higher in energy. This event was also observed by the machine learning based double pulse analysis \cite{icrcMLDP}. The event view of this event (Fig.~\ref{fig:2015event}) shows it is likely a background muon neutrino. First, the event starts inside the detector volume with a horizontal development direction, suggesting it is a neutrino event. Second, the horizontal development, seen especially on the leftmost strings in the event view, is an indication of a muon traveling out of the initial cascade and creating additional light as a track. The best fit event direction suggests that this muon leaves the detector volume soon after being produced, which creates an event that appears somewhat cascade like with a very short observable track. This is consistent with roughly one muon neutrino event expected in 8 years.

The final event was observed in 2017 and is two atmospheric muons that pass through the top of the detector one after the other, otherwise known as coincident muons. This event passed the selection due to a failure in the splitting algorithm used in IceCube. It is a known background but no representative simulated event passed to the final level. For these reasons, the event is removed from the event sample and not used for the forward folding analysis.

The forward folding was applied to the 2014 and 2015 event, excluding the 2017 event. The results of the fit for the three astrophysical spectra are shown in Table \ref{tab:folding}. The major take away from the fit results is the zero astrophysical tau neutrino flux normalization. The analysis prefers zero signal events and attributes the two observed events to background events. This is most likely due to no events in the region of a few 100 TeV where the majority of the signal events are expected. 

\begin{table}
  \centering
  \begin{tabular}{| l | c | c | r |}
  	  \hline
       & $E^{-2.19}$ & $E^{-2.5}$ & $E^{-2.9}$  \\ \hline
      $\nu_\tau$ Norm. &  0.0 & 0.0 & 0.0 \\ \hline
      $\nu_e$ Norm. &  0.0 & 1.76 & 0.0 \\ \hline
      $\nu_\mu$ Norm. &  0.0 & 1.93 & 1.81  \\ \hline
      Pi/K Ratio &  0.0 & 0.0 & 0.0 \\ \hline
      Conv. Norm. &  0.64 & 0.91 & 0.59 \\ \hline
      Prompt Norm. &  0.0 & 0.0 & 0.0  \\ \hline
      $\Delta \mathrm{CR}_\gamma$ &  -0.97 & -0.83 & -0.97  \\ \hline
  \end{tabular}
  \caption{The best fit values of the floating parameters for the three different assumed astrophysical flux spectra. All astrophysical normalizations are in units of $10^{-18} \mathrm{GeV}^{-1} \mathrm{cm}^{-2} \mathrm{s}^{-1} \mathrm{sr}^{-1}$.
  \label{tab:folding}}
\end{table}

\begin{figure}%
    \centering
    \includegraphics[width=0.50\textwidth]{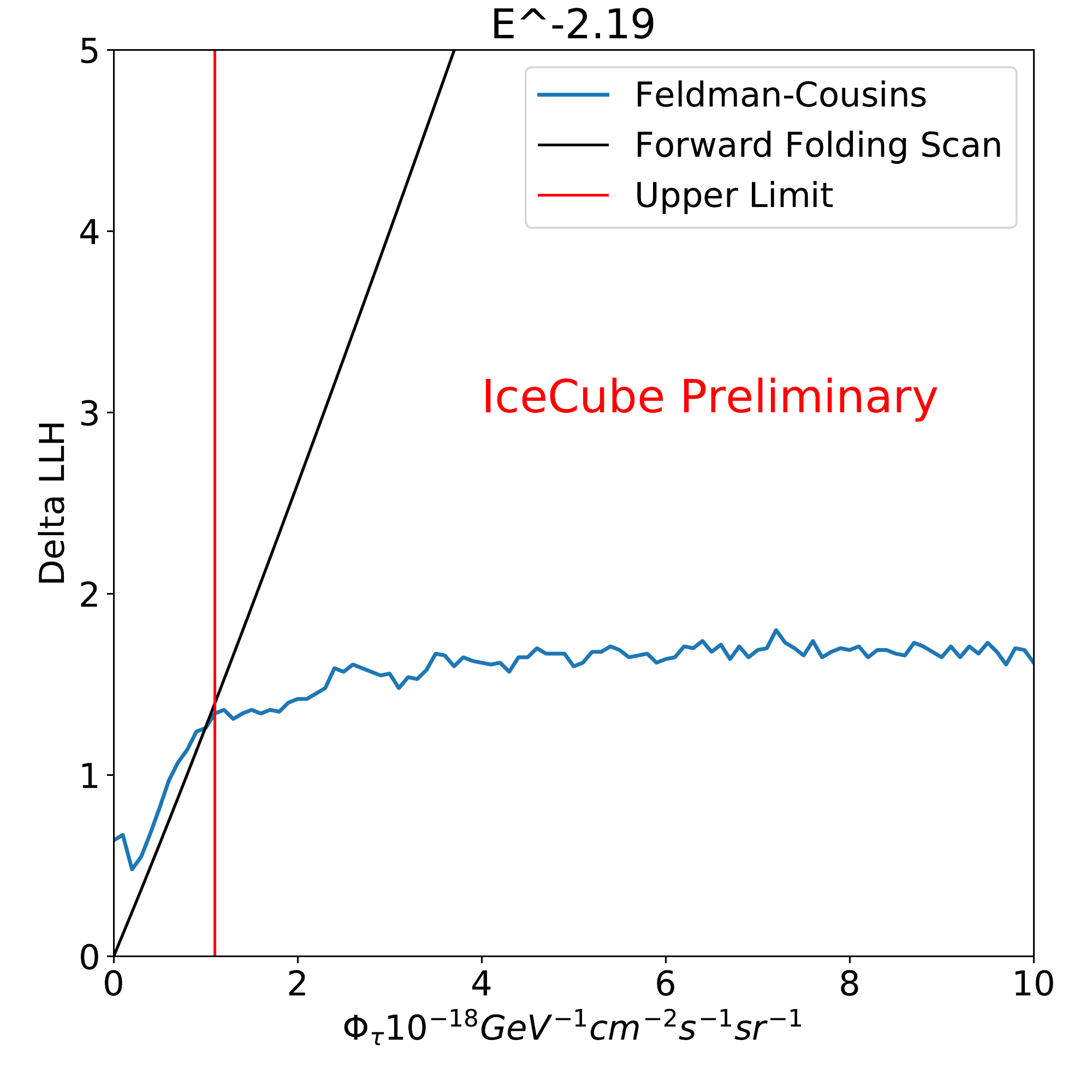}
    \caption{The likelihood scan of the tau neutrino astrophysical normalization plotted against a Feldman Cousins scan. The point at which the likelihood curve crosses the critical value determines the 90\% upper limit of the tau neutrino flux, this point is denoted with a vertical red line.}%
    \label{fig:FCscan}%
\end{figure}

The tau neutrino astrophysical flux normalization are scanned over taking the best fit values in Table \ref{tab:folding}. The likelihood difference from the best fit value for each normalization are shown in Fig. \ref{fig:FCscan}. A Feldman-Cousins' confidence interval method was applied to find the 90\% confidence upper limit of the tau neutrino astrophysical normalization \cite{Feldman:1997qc}. These upper limits are: $1.1 \times 10^{-18} \times \mathrm{E}^{-2.19} $, $2.5 \times 10^{-18} \times \mathrm{E}^{-2.5}$, and $6.0 \time 10^{-18} \times \mathrm{E}^{-2.9} \mathrm{GeV}^{-1} \mathrm{cm}^{-2} \mathrm{s}^{-1} \mathrm{sr}^{-1}$.

\section{Conclusion and Outlook}

The analysis method was applied to 2759.66 cumulative days of data taken by IceCube which had three events that passed the data selection process. The best fit flux from this analysis was zero tau neutrino astrophysical flux but the upper limit is not in conflict with previous astrophysical flux measurements. The sample of three events included one possible signal event, and one probable background event, and one certain background event. The event observed in 2014 had a slight indication of signal-like with p-values of 0.29, 0.196, and 1.0 for $E^{-2.19}$, $E^{-2.5}$, and $E^{-2.9}$ spectra respectively. The other two events have p-values of 1.0 for all spectra and their event views show topologies of background events. While the 2014 event is inconclusive if it is a tau neutrino event, a posteriori analyses are ongoing to further analyze this event. Three upper limits were constructed, $1.1 \times 10^{-18} \times \mathrm{E}^{-2.19} $, $2.5 \times 10^{-18} \times \mathrm{E}^{-2.5}$, and $6.0 \time 10^{-18} \times \mathrm{E}^{-2.9} \mathrm{GeV}^{-1} \mathrm{cm}^{-2} \mathrm{s}^{-1} \mathrm{sr}^{-1}$. These are not in conflict with a 1:1:1 flavor ratio, as the measured normalization for these three spectra are below the upper limits.

\bibliographystyle{ICRC}
\bibliography{references}

\end{document}